\documentclass[useAMS,usenatbib,usegraphicx]{mn2e}  
\usepackage[usenames]{color}

\newcommand{\be}{\begin{equation}}
\newcommand{\ee}{\end{equation}}
\newcommand{\bea}{\begin{eqnarray}}
\newcommand{\eea}{\end{eqnarray}}

 \title[DM in MACS1149] 
{A Free-Form Prediction for the Reappearance of Supernova Refsdal in the Hubble Frontier Fields Cluster MACSJ1149.5+2223.}
\author[J.M. Diego]  
  {Jose M. Diego\thanks{jdiego@ifca.unican.es}$^1$, Tom Broadhurst$^{2,3}$, Cuncheng Chen$^6$, Jeremy Lim$^6$, Adi Zitrin$^{4,5}$, 
  \newauthor
  Brian Chan$^6$, Dan Coe$^7$, Holland C. Ford$^7$, Daniel Lam$^6$,Wei Zheng$^7$\\
$^{1}$IFCA, Instituto de F\'isica de Cantabria (UC-CSIC), Av. de Los Castros s/n, 39005 Santander, Spain\\
$^{2}$Fisika Teorikoa, Zientzia eta Teknologia Fakultatea, Euskal Herriko 
Unibertsitatea UPV/EHU\\ 
$^3$IKERBASQUE, Basque Foundation for Science, Alameda Urquijo, 36-5 48008 Bilbao, Spain\\
$^4$ Cahill Center for Astronomy and Astrophysics, California Institute of Technology, MS 249-17, Pasadena, CA 91125, USA.\\
$^5$ Hubble Fellow\\
$^6$ Department of Physics, The University of Hong Kong, Pokfulam Road, Hong Kong \\
$^7$ Dept. of Physics and Astronomy, Johns Hopkins University, Baltimore, Maryland, USA
}

\date{Draft version \today}  
  
\pagerange{\pageref{firstpage}--\pageref{lastpage}}  
  
\begin{document}  
\maketitle  
 
\label{firstpage}  
\begin{abstract} 

The massive cluster MACSJ1149.5+2223(z=0.544) displays five very large lensed images of a well resolved spiral galaxy at $z_{\rm spect}=1.491$. It is within one of these images that the first example of a multiply-lensed  supernova has been detected recently as part of the Grism Lens-Amplified Survey from Space. The depth of this data also reveals many HII regions within the lensed spiral galaxy which we identify between the five counter-images. Here we expand the capability of our free-form method to incorporate these HII regions locally, with other reliable lensed galaxies added for a global solution. This improved accuracy allows us to estimate when the Refsdal supernova will appear within the other lensed images of the spiral galaxy to an accuracy of $\sim$ 7\%. We predict this supernova will reappear in one of the counter-images (RA=11:49:36.025, DEC=+22:23:48.11, J2000) and on November 1$^{st}$ 2015 (with an estimated error of $\pm$25 days) it will be at the same phase as it was when it was originally discovered, offering a unique opportunity to study the early phases of this supernova and to examine the consistency of the mass model and the cosmological model that have an impact on the time delay prediction.

\end{abstract}  
  
\section{Introduction}\label{sect_intro}  
The unprecedented data quality of the Hubble Frontier Fields (HFF) 
program\footnote{http://www.stsci.edu/hst/campaigns/frontier-fields/} \citep{Lotz2014}
provides a good opportunity to study the mass distribution in the central region of merging clusters in detail. In addition, the HFF data is very useful to study the population of high-z galaxies and to discover new supernovas (or SN hereafter). One of the this SN, \citep[SN Refsdal,][]{Kelly2015} is one of the main subjects of this paper together with the mass distribution in the core of MACS1149\footnote{{\bf While this article was in press, news of the 
reappearance of SN Refsdal broke on December 12 2015, confirming the prediction of this paper.}} 
.
   
The HFF images contain many tens of multiply lensed images that are not easily recognised, but require the guidance of a reliable model. This is partially due to the complexity of the clusters chosen for the HFF program that maximises the lensing signal. During a major merger the critical curves can be {\it stretched} between the mass components enhancing the critical area with elongated critical curves \citep{Zitrin2013}. This effect results in a relatively large sky area subjected to very large magnification and hence to the detection of unusually bright lensed galaxies. Galaxies as distant as z$\simeq$ 10 have been identified through the HFF program \citep{Zitrin2014,Zheng2014,Oesch2014,Coe2015,Ishigaki2015} with the potential to reach z$\simeq$ 12 given the spectral coverage of the HFF data. 

The HFF program may in fact be probing already the edge of the observable universe at near-IR wavelength as so far there as yet no examples of galaxies beyond z$\simeq$ 10 in the HFF data. This lack of higher redshift galaxies may  be supported by the  recently updated value for the mean redshift of reionization calculated from from the inferred value of the optical depth, $\tau$ obtained by the {\it Planck} mission data for which a lower redshift (compared with previous estimations based on CMB data) of z $\simeq$ 8.8 (for instantaneous or mean redshift of reionization) has been estimated, \citep{PlanckCosmo2015}. This has implications for the assumptions regarding the spectral index of the ionising radiation,
and the extrapolation of the galaxy UV luminosity function to undetected luminosities together with the escape fraction of ionising radiation from high-z galaxies. Despite these considerable uncertainties, consistency has been claimed between the recent low value of $\tau$  from Plank and the rough first measurements of the UV selected luminosity density of z$>$9 galaxies \citep{Robertson2015,Bouwens2015}.  The initially claimed {\it steep decline} in the integrated UV luminosity density of galaxies at z$>$9  would seem to empirically support this z$\sim$9 epoch as marking the beginning of galaxy formation \citep{Oesch2012,Oesch2014} and interestingly this is not obviously reconciled with the many predictions made for $\lambda$CDM with ever smaller galaxies naturally expected  to higher redshift, in a scale free way, limited only by the relatively large Jeans scale for metal free star formation \citep{McKee2007,Barkana2006}. A lower redshift of galaxy formation is anticipated for CDM in the form of  light bosons limited by a Jeans scale for the Dark Matter generated by quantum pressure of bosons in the ground state \citep{Peebles2000,Hu2000}. The first simulations of this form of CDM normalised to fit local galaxy dark matter (or DM) cores predict the first galaxies at z$\simeq$ 12 in a 30 Mpc volume \citep{Schive2014} and hence this interpretation of CDM is more viable than heavy fermionic WIMPs that are increasingly undetected in the laboratory.  The HFF may provide considerably more clarity in deciding between these two very different interpretations of CDM by providing sufficient z$>$9 galaxies with lower luminosities than field surveys, by virtue of the high levels of lens magnification.

\begin{figure}    
 {\includegraphics[width=8.0cm]{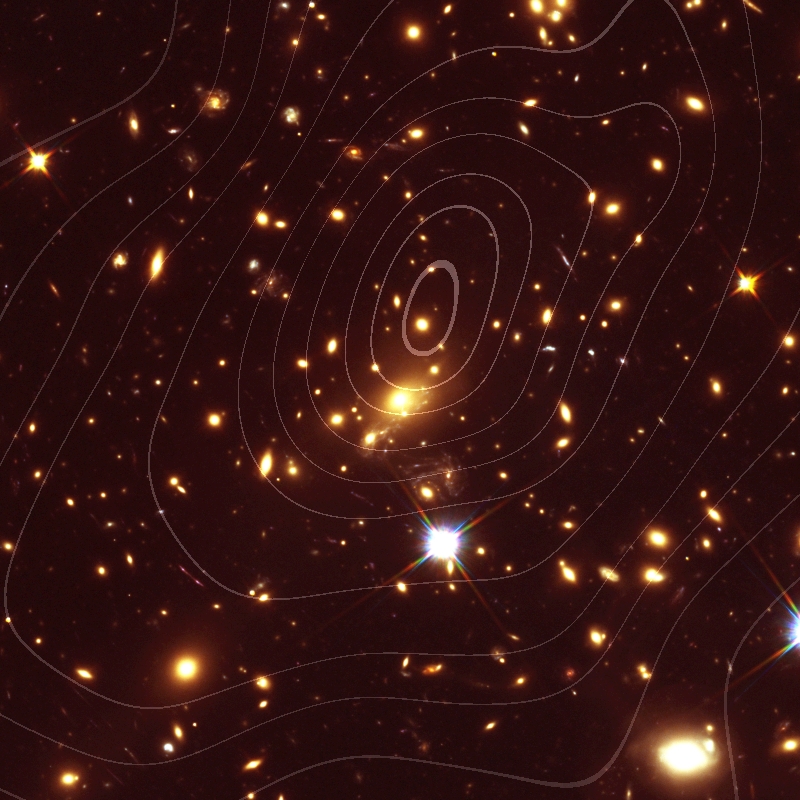}}
   \caption{MACS1149 as seen by HST with {\it Chandra} contours overlaid on top. The field of view is 1.6\arcmin.}
   \label{fig_MACS1149_Chandra}  
\end{figure}  

Another important reason to study the HFF clusters is for the constraints that may be derived from the level of any self-interaction within the dark matter, as those clusters are caught in the act of collision. Although relaxed clusters seem to agree well with predictions from  $\lambda$CDM models \citep{Newman2013}, recent results on non-relaxed clusters show interesting deviations in the density profiles in the central regions when compared to predictions from standard $\lambda$CDM.  In particular, shallow profiles have been identified by several authors in HFF clusters and more in agreement with self-interacting dark matter expected for purely collisionless dark matter \citep{Diego2014,Lam2014, Diego2015}.  Other possible, and less exotic, interpretations may be related to projection effects, overlapping of cluster cores and uncertainties in the reconstruction of the central regions.  More data and better hydrodynamical modeling of the HFF clusters may help resolve these questions.  It is imperative therefore that  free-form modeling of these lenses  is achieved to reliably establish the density of background sources detected as a function of redshift.

In this paper  we explore the remarkable MACS1149(z=0.544) for which 5 very large well resolved images of a spiral galaxy were recognised as 
multiple images \citep{Zitrin2009a,Smith2009} and subsequently explored more thoroughly with the CLASH program \citep{Rau2014,Zitrin2015}. These models demonstrate that the magnification is large over the full critical area of this cluster because of the flatness of the central density profile which lies close to the critical value for lensing in the central strong lensing region and is hence optimal for creating high magnification. Indeed a very high redshift galaxy has been identified by the CLASH team \citep{Zheng2012} at z$\simeq$ 9.6, which has a high magnification. The power of this lens has led to its selection for the HFF program in the search for more higher redshift lensed galaxies. The depth of the HFF data allows now  many internal HII regions within 5 spiral galaxy images to be identified and matched between the counter images. Moreover, the distribution of the multiple counterimages map the central region in a semi-continuous fashion from distances $R=13.12\arcsec$ (or $R=84.62$ kpc) to only 
$R=1.26\arcsec$ (or $R=8.12$ kpc) from the centre of the dominant BCG. In addition, a SN is observed multiply lensed 4 times around a cluster member galaxy \citep{Kelly2015}. This SN is observed in only one of the counterimages to date. The corresponding SN event in the other counter images has either occurred in the past of will happen in the future. Time delays 
between the different counterimages have been computed by different authors but with estimates varying in general by several years between the different authors \citep{Oguri2015,Sharon2015,Kelly2015}. The time delay between the counterimages depend on the lens model, position of the background source and cosmology (in particular the Hubble constant).  Having an accurate time delay for this SN will be useful to plan an observing campaign of the {\it future} SN. This will offer the rare opportunity to study the SN from the very early phases, providing also a test of competing lensing models and a consistency check of cosmological models.

Here we implement an enhancement to our free-form method, WSLAP+  \citep{Diego2005,Diego2007,Sendra2014}, which is largely motivated by the uniqueness of the particularly large lensed images generated by  MACSJ1149 ($z=0.544$) and we calculate the corresponding time delays for the SN event. The paper is organized as follows. We describe the Hubble data in Section \ref{sect_S2}.
The X-ray data is described in Section \ref{sect_S3}. 
The lensing data is described in Section \ref{sect_S4}.
In Section \ref{sect_S5} we give a brief description of the reconstruction method with the new improvements that are applied to the data for the first time in this work. 
Section \ref{sect_S6} describes six different scenarios that are assumed to reconstruct the lens and to study the uncertainties and variability in the solutions. 
Section \ref{sect_S7} presents the results of the lensing analysis, focusing on the reproducibility of system 1, the 2-dimensional mass distribution, the projected mass profile 
and the time delays for the Refsdal SN. 
We conclude in Section \ref{sect_S8}.

Throughout the paper we assume a cosmological model with $\Omega_M=0.3$,
$\Lambda=0.7$, $h=70$ km/s/Mpc. For this model, 1\arcsec
equals $6.45$ kpc at the distance of the cluster.

\section{HFF data}\label{sect_S2}
In this paper we use public imaging data obtained from the ACS (filters: F435W, F606W and F814W) 
and the WFC3 (F105W, F125W, F140W and F160W), retrieved from the Mikulski Archive for Space 
Telescope (MAST). The data used in this paper consists of $\approx 1/3$ of the data to be collected 
as part of the HFF program. 
Part of the data comes from CLASH \citep{Postman2012}. This release includes the first 70 orbits of
observations of MACS1149 from the Frontier Fields program ID 13504 (PI. J. Lotz)
but also including archival ACS and WFC3/IR data from programs 13790 (PI. S. Rodney), and 14041 (PI. P. Kelly). 
In the IR bands we use the background corrected images, corrected for a time-dependent increase in the background sky level \citep[see for instance][]{Koekemoer2013}. 
From the original files, we produce two sets of color images combining the optical and IR bands. The first set is based on the raw data while in the second set we apply a low-pass filter to reduce the diffuse emission from member galaxies and a high-pass filter to increase the signal-to-noise ratio of small compact faint objects. The second set is particularly useful to match colors in objects that lie behind a luminous 
member galaxy. 

\section{X-ray data}\label{sect_S3}
\begin{figure}    
 {\includegraphics[width=9.0cm]{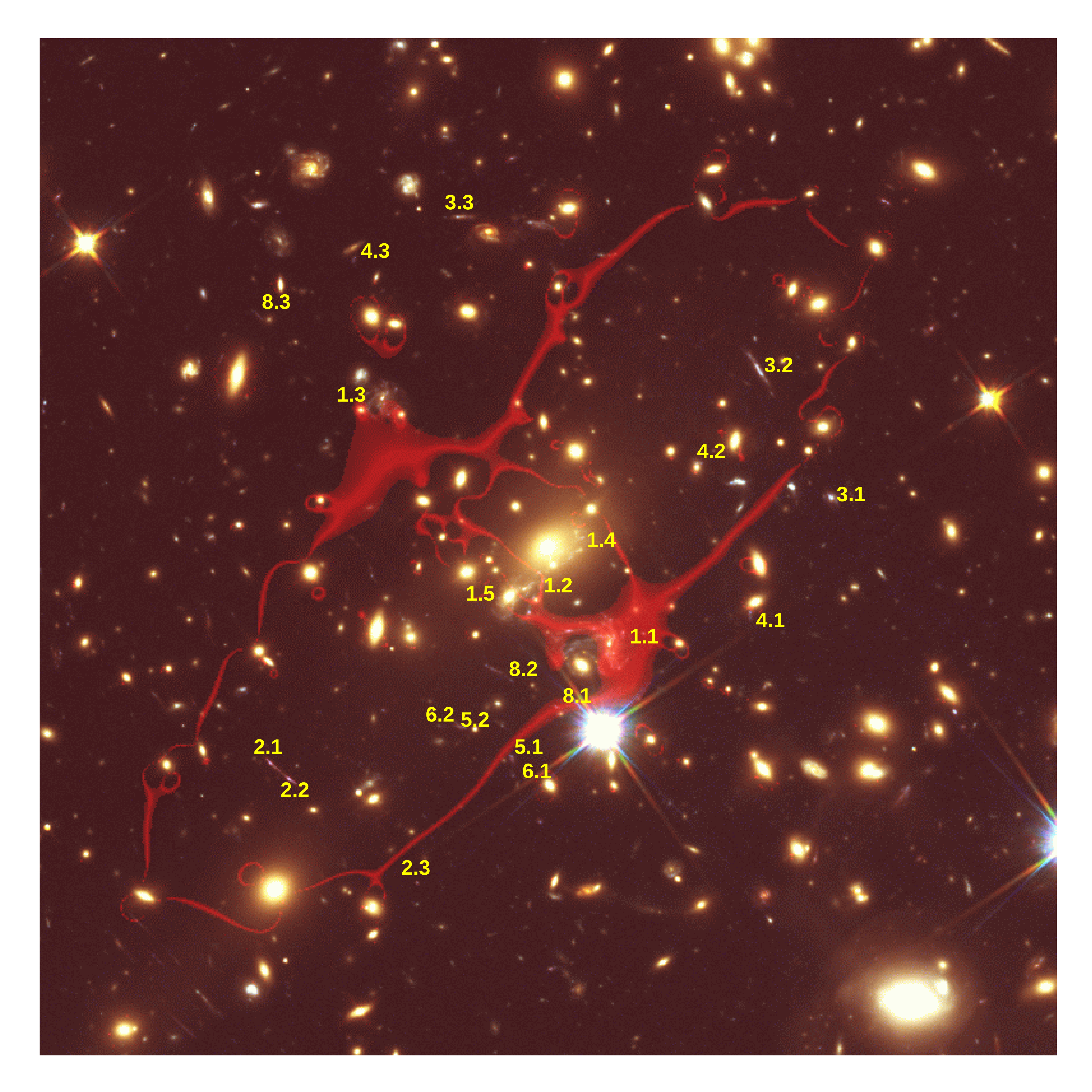}}
   \caption{MACS1149 with a typical critical curve ($z_{\rm s}=3$) for one of our models (case 5, see text). 
            The images used in the reconstruction are marked with yellow IDs. 
            The field of view is 1.6\arcmin.}
   \label{fig_MACS1149_CritCurve}  
\end{figure}  

To explore the dynamical state of this cluster, we produce an X-ray image using recent public Chandra data on this cluster. 
In particular we use data with the following Obs IDs, 1656, 3589, 16238, 16239, 17595, 17596, 16306, 16582 (PIs. Vaspeybroeck, Jones, Murray) totaling 363.4 ks. 
The X-ray data is smoothed using the code {\small ASMOOTH} \citep{Ebeling2006}. 
The smoothed X-ray map is compared with the distribution of galaxies in figure \ref{fig_MACS1149_Chandra}. 

A significant  offset is observed between the peak of the X-ray emission and the BCG indicating that this 
cluster shows the effect of collision. The X-ray emission is elongated in the diagonal direction and, as discussed later, the same elongation is found in the distribution of matter although the peak of the mass distribution is also found to be offset 
with respect to the peak of the X-rays and more in agreement with the position of the BCG.

\section{Lensing data}\label{sect_S4}

\begin{figure*}  
   \includegraphics[width=8.5cm]{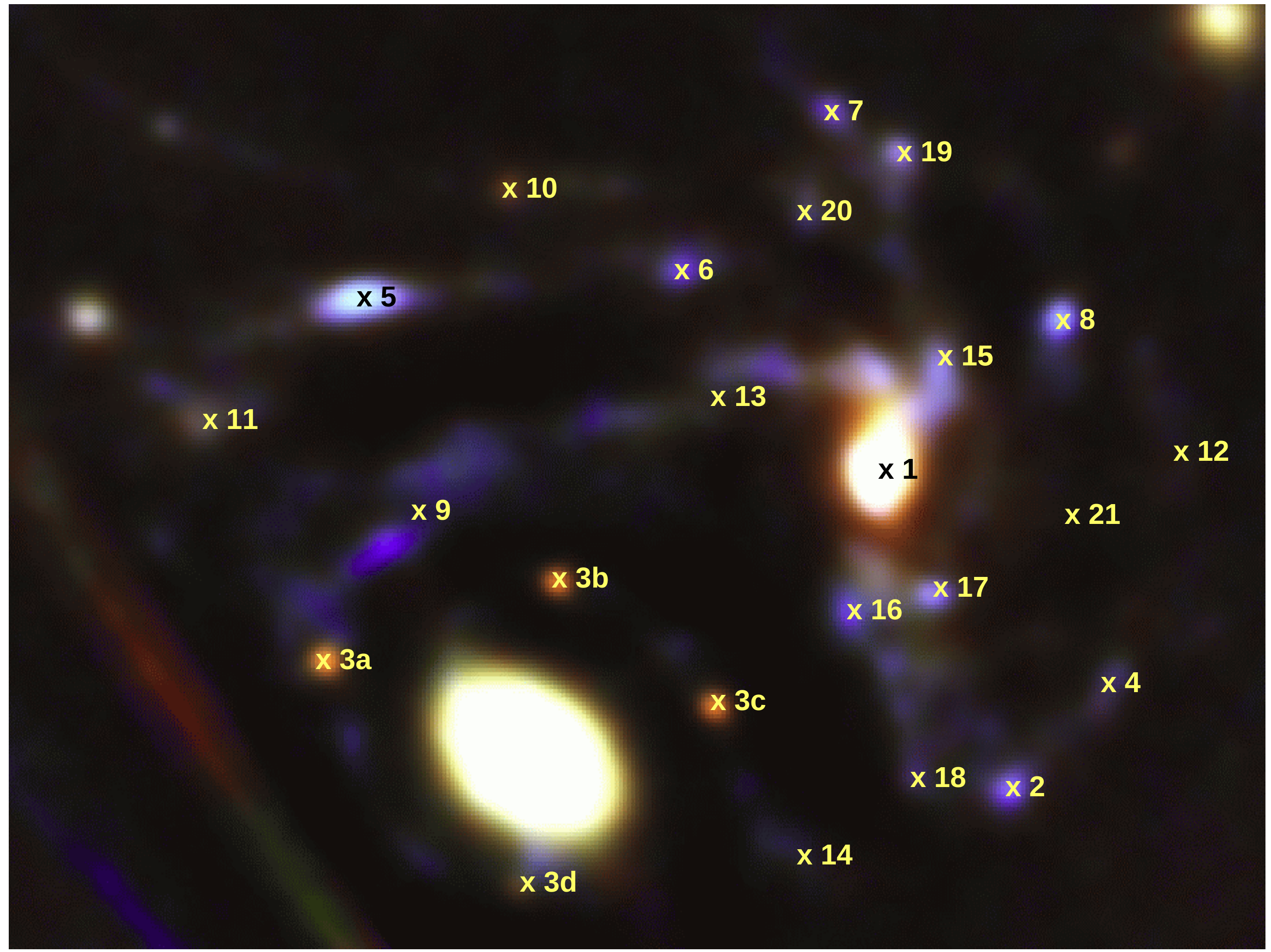} 
   \includegraphics[width=7.3cm]{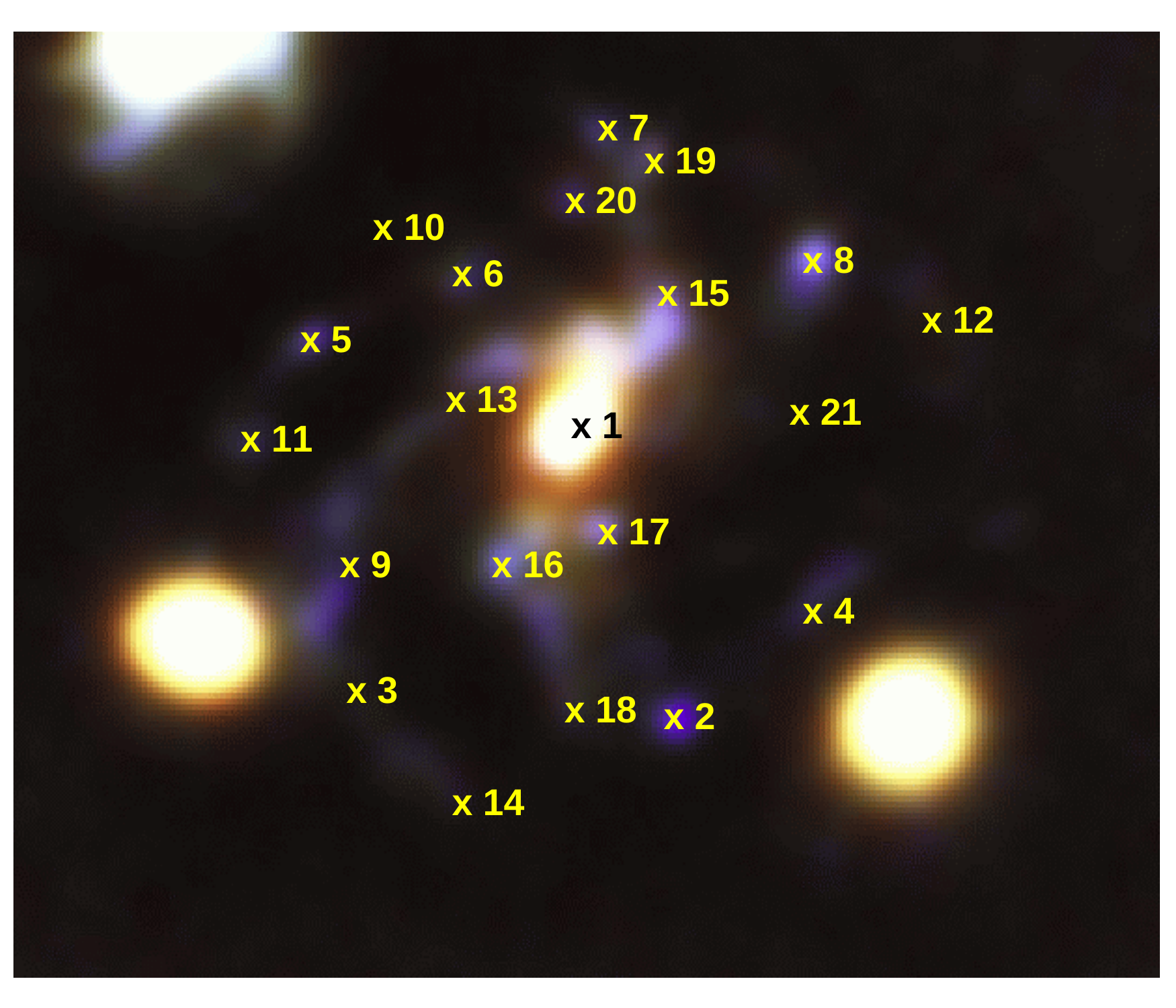} 
   \caption{Knots for system 1 used to do the mass reconstruction. Only two of the counterimages of system 1 are represented here.} 
    \label{fig_systemID}  
\end{figure*}  

For the lensing data we adopt the previous multiple-image system identifications from CLASH data \citep{Zitrin2015} from which we also adopt his numbering system except for systems 2 and 3 that are swapped like in \cite{Smith2009}.
In particular, we use the reliable  systems 1,2,3,4,5,6 and 8. For systems 5 and 6 we use only two of the counterimages as the remaining third counterimage is not regarded as reliable (various candidates are consistent with being the counterpart of the third images, 5.3 and 6.3). Further HFF data will soon help clarify these and uncover other systems. Systems 1,2, and 3 have spectroscopic redshifts that we take from the compilation of \cite{Zitrin2015} and were originally measured by \cite{Smith2009}. The redshifts of systems 4,5,6 and 8 are matched to the redshifts preferred by the lens models (see below) which constrain the position of the critical curve.   For these four systems, the redshifts preferred by our lens model are between 10\% and 30\% higher than the inferred redshifts (also from a lens model) derived by \cite{Zitrin2015}. We should note that the redshifts for systems  4,5,6 and 8 may differ from estimates used also by other authors. This may result in differences in the derived mass model. More specifically, for systems 1,2,3,4,5,6 and 8 we adopt the following redshifts; 1.491, 1.894, 2.497, 2.5, 1.9, 1.9, and 2.5 respectively.  The systems are shown in Fig. \ref{fig_MACS1149_CritCurve}. For comparison purposes we also show the critical curve for one of our models for a source  at redshift $z=3$.

In addition to the position of these lensed systems, for some of them we use the position of knots readily identified in the different 
counterimages thanks to the depth of the data (see Fig. \ref{fig_systemID}). The large image 1.1 \citep[in the notation of][] {Zitrin2009a} is the largest, most magnified image of system 1 (and displays the 4 SN images) with many internal features that we label in Figure \ref{fig_systemID}.
Lensed systems 3 and 8 are also morphologically resolved allowing us to identify a few individual knots and these are also included in our new reconstruction algorithm.   In order to take full advantage of the resolved geometry of system 1 (and to a lesser extent of systems 3 and 8), we introduce two enhancements to our code that are described in the next section.

\section{Lensing reconstruction and improvements to the code}\label{sect_S5}
We use the method, WSLAP+, that we have been developing to perform the lensing mass reconstruction
with the lensed systems and internal features described above.
The reader can find the details of the method in our previous papers
\citep{Diego2005,Diego2007,Sendra2014}. 
Here we give a brief summary of the most essential elements. \\
Given the standard lens equation, 
\begin{equation} \beta = \theta -
\alpha(\theta,\Sigma), 
\label{eq_lens} 
\end{equation} 
where $\theta$ is the observed position of the source, $\alpha$ is the
deflection angle, $\Sigma(\theta)$ is the surface mass density of the
cluster at the position $\theta$, and $\beta$ is the position of
the background source. Both the strong lensing and weak lensing
observables can be expressed in terms of derivatives of the lensing
potential. 
\begin{equation}
\label{2-dim_potential} 
\psi(\theta) = \frac{4 G D_{l}D_{ls}}{c^2 D_{s}} \int d^2\theta'
\Sigma(\theta')ln(|\theta - \theta'|), \label{eq_psi} 
\end{equation}
where $D_l$, $D_s$, and $D_{ls}$ are the
angular diameter distances to the lens, to the source and from the lens to 
the source, respectively. The unknowns of the lensing
problem are in general the surface mass density and the positions of
the background sources in the source plane. 
The surface mass density is described by the combination of two components; \\
$\bullet$ i) a soft (or diffuse) component that is parametrised as a superposition of Gaussians on a grid of constant 
width (regular grid) or varying width (adaptive grid). \\
$\bullet$ ii) a compact component that accounts for the mass associated with the individual halos (galaxies) in the cluster. This component is modelled either as NFW profiles with a mass proportional to the light of each galaxy 
or adopting directly the light profile (in one of the IR bands). 

The compact component is usually divided 
in independent layers, each one containing one or several cluster members. The separation in different layers allows us to constrain the mass associated to special halos (like the ones from giant elliptical galaxies) independently from more ordinary galaxies. This is useful in the case where the light-to-mass ratio may be different, like for instance in the BCG. \\

As shown in \cite{Diego2005,Diego2007}, the strong and weak lensing problem can be expressed as a system of linear
equations that can be represented in a compact form, 
\begin{equation}
\Theta = \Gamma X, 
\label{eq_lens_system} 
\end{equation} 
where the measured strong lensing observables (and weak lensing if available) are contained in the
array $\Theta$ of dimension $N_{\Theta }=2N_{SL}$, the
unknown surface mass density and source positions are in the array $X$
of dimension $N_X=N_c + N_g + 2N_s$ and the matrix $\Gamma$ is known
(for a given grid configuration and fiducial galaxy deflection field) 
and has dimension $N_{\Theta }\times N_X$.  $N_{SL}$ is the number
of strong lensing observables (each one contributing with two constraints,
$x$, and $y$), $N_c$ is the number of grid points (or cells) that we use to divide
the field of view. 
Each grid point contains a Gaussian function. The width of the Gaussians are chosen in such a way 
that two neighbouring grid points with the same amplitude produce a horizontal plateau in between the two 
overlapping Gaussians.
$N_g$ is the number of deflection fields (from cluster members) that we consider.  
In this work we set $N_g$ equal to 3. The first deflection field contains the BCG galaxy, the second 
contains a prominent elliptical galaxy near the image 1.2 and the third deflection field contains the remaining 
galaxies (N=121) from the cluster that are selected from the red-sequence 
(totaling 123 galaxies between the 3 layers). Dividing the cluster galaxies in 3 layers allows us to 
independently fit the mass of the two giant ellipticals from the other galaxies. The particular configuration of the galaxies 
in the three layers is shown in figure \ref{fig_MassChandra}. 
$N_s$ is the number of background sources (each contributes with two unknowns, 
$\beta_x$, and $\beta_y$) which in our particular case is $N_s=7$ . The solution is found after
minimising a quadratic function that estimates the solution of the
system of equations (\ref{eq_lens_system}). For this minimisation we
use a quadratic algorithm which is optimised for solutions with the
constraint that the solution, $X$, must be positive. Since the vector $X$ contains the grid masses, 
the re-normalisation factors for the galaxy deflection field and the background source positions, and all these 
quantities are always positive (the zero of the source positions is defined in the bottom left corner of the 
field of view), imposing  $X>0$ helps in constraining the space of meaningful solutions. 
The condition $X>0$ also helps in regularising the solution as it avoids large negative and positive 
contiguous fluctuations. The quadratic algorithm convergence is fast (few minutes) on a desktop allowing 
for multiple solutions to be explored on a relatively sort time. Different solutions can be obtained after modifying 
the starting point (or grid configuration) in the optimization. A detailed discussion of the quadratic algorithm can be found 
in \cite{Diego2005}. A recent discussion about its convergence and performance can be found in \cite{Sendra2014}.

\begin{figure}  
   \includegraphics[width=8cm]{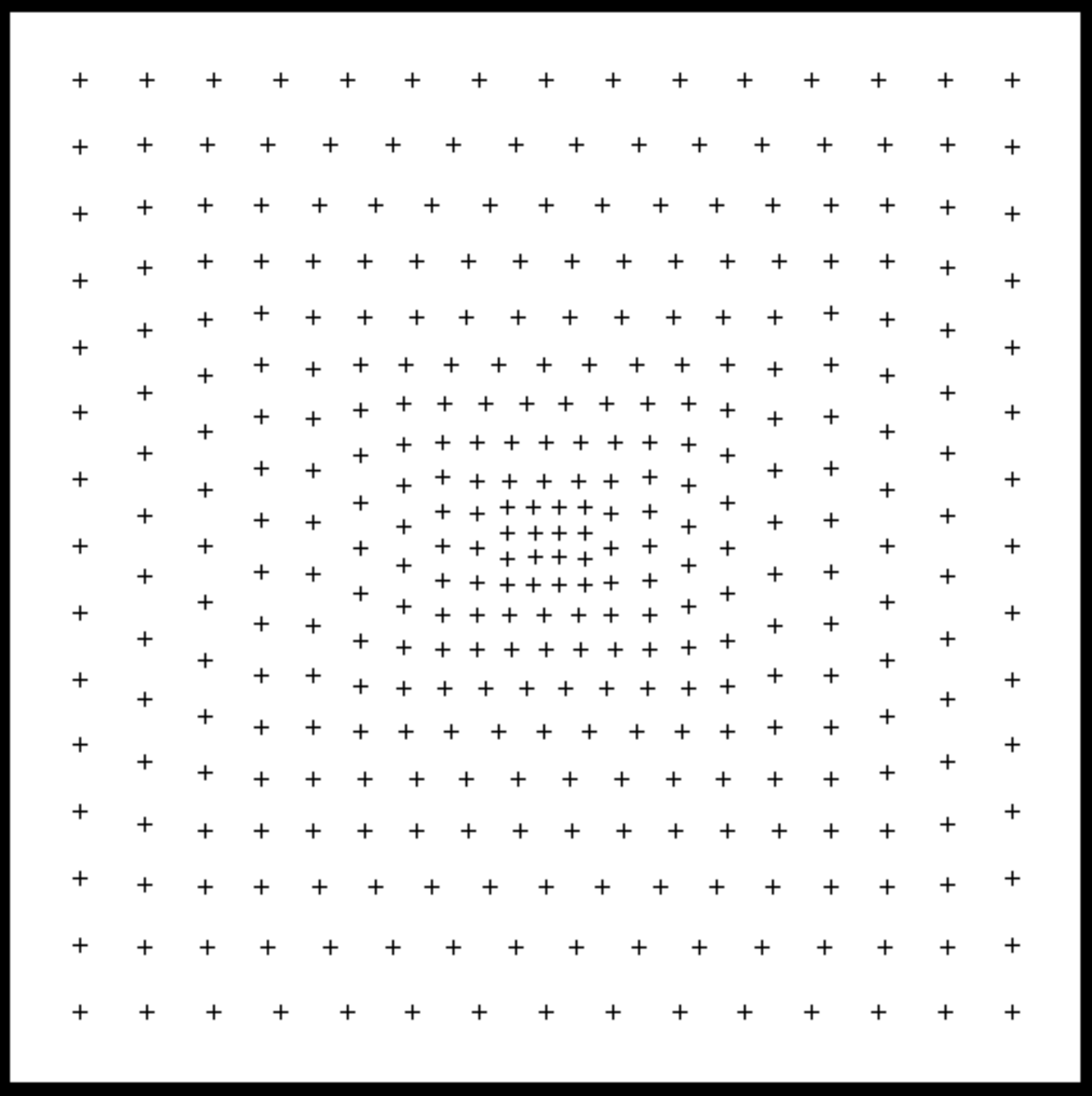} 
   \caption{Example of a multiresolution grid (352 grid points). The peaks of the individual 
   Gaussians are located at the positions of the crosses. The width of  the Gaussians are adjusted to 
   guarantee a smooth constant distribution when the amplitudes are equal. This distribution is suited to the 
   increasing strong lensing data density towards the center. } 
    \label{fig_multires_grid}  
\end{figure}

\subsection{New Enhancements}
Two improvements are implemented that resolve some of the issues found in the previous version of the code. One of the problems of the original code was 
a systematic bias introduced in the reconstructed solution when using a multiresolution grid. This bias was the consequence of sharp changes in resolution between 
neighbouring cells (or Gaussians). The difference in size between two cells of different resolutions was a factor $2^n$ which introduced spurious artifacts in the space 
dividing two resolutions. We have mitigated this problem by introducing a more gradual change 
between neighbouring cells. An example of the new scheme for the multiresolution grid is shown in figure \ref{fig_multires_grid}. The use of the multiresolution 
grid with  increased resolution around the BCG has the advantage of allowing for a more complex mass distribution in the central region where the density of lensed image constraints is higher  and in particular it allows 
for a better and more flexible way of parametrising the elongation of the dark matter halo in the central region. As a general rule, when using the new multiresolution grid we always increase the density of grid points towards the centre (BCG). The difference in the width of the Gaussians between two consecutive grid points is a constant fraction that can be changed depending on the desired degree of freedom. 

A second improvement is related with the original assumption that the sources are very compact. This assumption is normally a good approximation 
but in cases like system 1, this assumption would result in unphysical solutions that predict a very small source for system 1 in the source plane. This pathological problem 
was extensively discussed in our earlier work \citep{Diego2005,Diego2007,Ponente2011} and the solutions derived from it were referred to as the {\it point source} solution.
Information related to the shape of the galaxies in the source plane can be easily integrated in the algorithm. 
Based on a preliminary solution that avoids this pathological behaviour it is possible to produce a good guess 
for the shape and size of the galaxy in the source plane. 
This information on the expected shape (and size) of the source can be incorporated in the algorithm as additional constraints. 
The minimization process then converges to a solution that is 
stable and does not produce unphysical solutions (like the {\it point source} solution discussed in our earlier work). 
A similar behaviour was observed when introducing the deflection field from the member galaxies as part of the lens model as they act as an 
{\it anchor} for the solutions, better constraining the range of possible solutions. 
A future improvement of the code will include a penalty function for those models that predict images (knots) at positions that are  not observed. 
This approach was already initially explored in \cite{Diego2005} with promising results and referred to as the {\it null space}). 

\begin{figure}  
   \includegraphics[width=8.5cm]{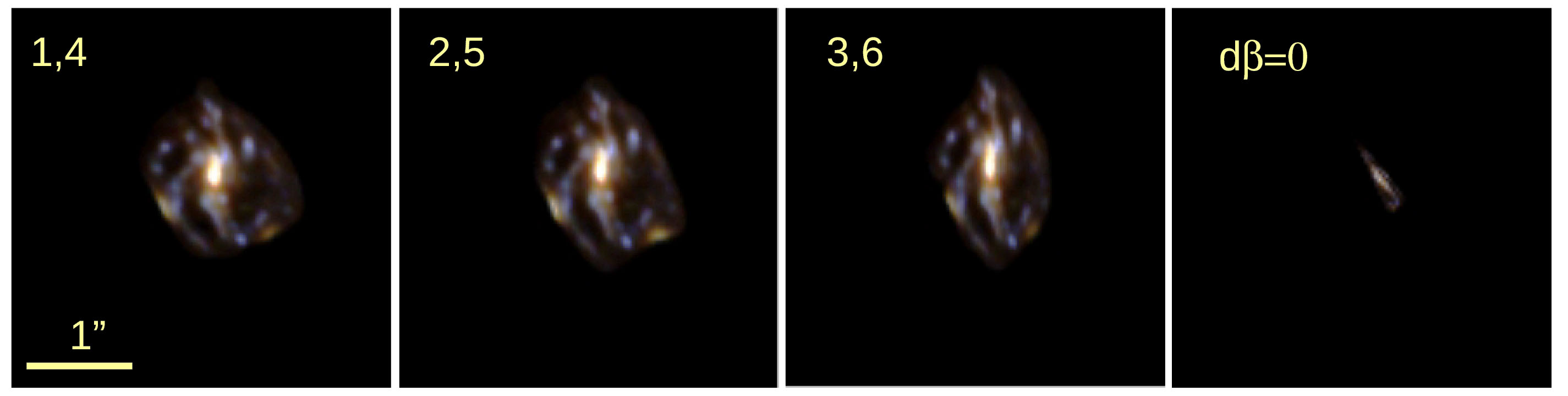} 
   \caption{Delensed image 1.3 for the 6 models.
            Only the cases for models 1,2 and 3 are shown. Cases 4,5 and 6 are virtually indistinguishable 
                 from 1,2 and 3 respectively. 
            For comparison, the panel on the right end shows the case for the same number of iterations but without 
            a constrain on the size of the source (or singular case). The number indicates the case discussed in section \ref{sect_S6}. 
            All figures have the same scale.}
    \label{fig_delened1c}  
\end{figure}  

\begin{figure}  
   \includegraphics[width=8.5cm]{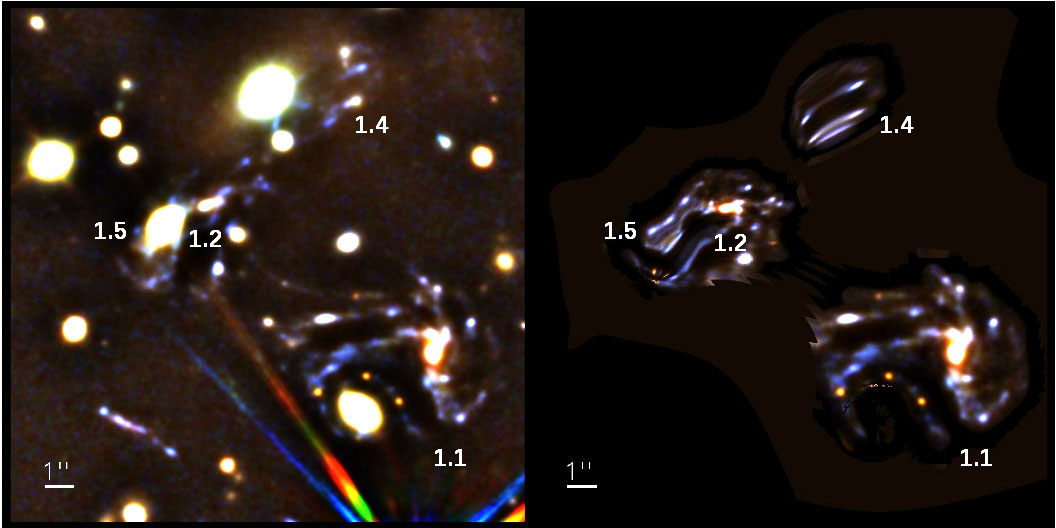} 
   \caption{Original (left panel) versus model-predicted (right panel) counterimages for system 1 (the model corresponds to case 5). 
            The largest counterimage 1.1 (bottom-right of the left panel) is used to predict the other counterimages. 
            The centre and field of view is the same in both panels. Note how the reproduction of 1.4 is still affected 
            by significant errors probably related to imperfect modeling of the BCG and neighbouring galaxy at $\approx 1\arcsec$ 
            south from the BCG.}
    \label{fig_Relens1a}  
\end{figure}

\subsection{Models}\label{sect_S6}
To account for uncertainties and variability in the solutions, we explore a range of cases where we change the assumptions for the two main components 
of our method: the member galaxies and the grid definition. In particular we consider six types of models (or cases) described briefly below.\\
 
$\bullet$ Case 1. We use a standard grid of $16\times16=256$ cells in our field of view. Each member galaxy is assigned an NFW profile where its total mass 
is taken proportional to its luminosity in the 814W filter band. The scale radius of the NFW is derived from the mass assuming a scaling $M^{1/3}$. 
The concentration parameter is fixed to $C=8$.\\
$\bullet$ Case 2. Like Case 1 but instead of a uniform regular grid we use a multiresolution grid with 280 cells similar to the one  shown in figure \ref{fig_multires_grid}.\\
$\bullet$ Case 3. Like Case 2 but instead of a multiresolution grid with 280 cells we increase the resolution and use a grid with 576 cells. \\
$\bullet$ Case 4. Like Case 1 but we divide the scale radius by a factor 2 making the galaxies more compact.\\
$\bullet$ Case 5. Like Case 2 but we divide the scale radius by a factor 2 making the galaxies more compact.\\
$\bullet$ Case 6. Like Case 3 but we divide the scale radius by a factor 2 making the galaxies more compact.\\

In addition to these cases, we briefly consider the equivalent of Case 1 but in our previous implementation of the code where no information about the spatial extent of system 1 
is used and the galaxy in the source plane is assumed to be very compact. We refer to this case as the {\it singular} case and would correspond to the solution obtained with our previous version of the code in the situation of maximum convergence \citep[referred as {\it point source} solution in][]{Diego2005,Sendra2014}. 
In all cases, we assume three deflection fields for the galaxies as described in the previous section. 
Most of the galaxies are contained in one deflection field. The BCG and the elliptical next to image 1.2 are treated 
as independent deflection fields and their masses are re-scaled by the algorithm in the minimization process. 
For the remaining cluster members, their masses are also re-scaled but all by the same factor.

\begin{figure*}  
   \includegraphics[width=16cm]{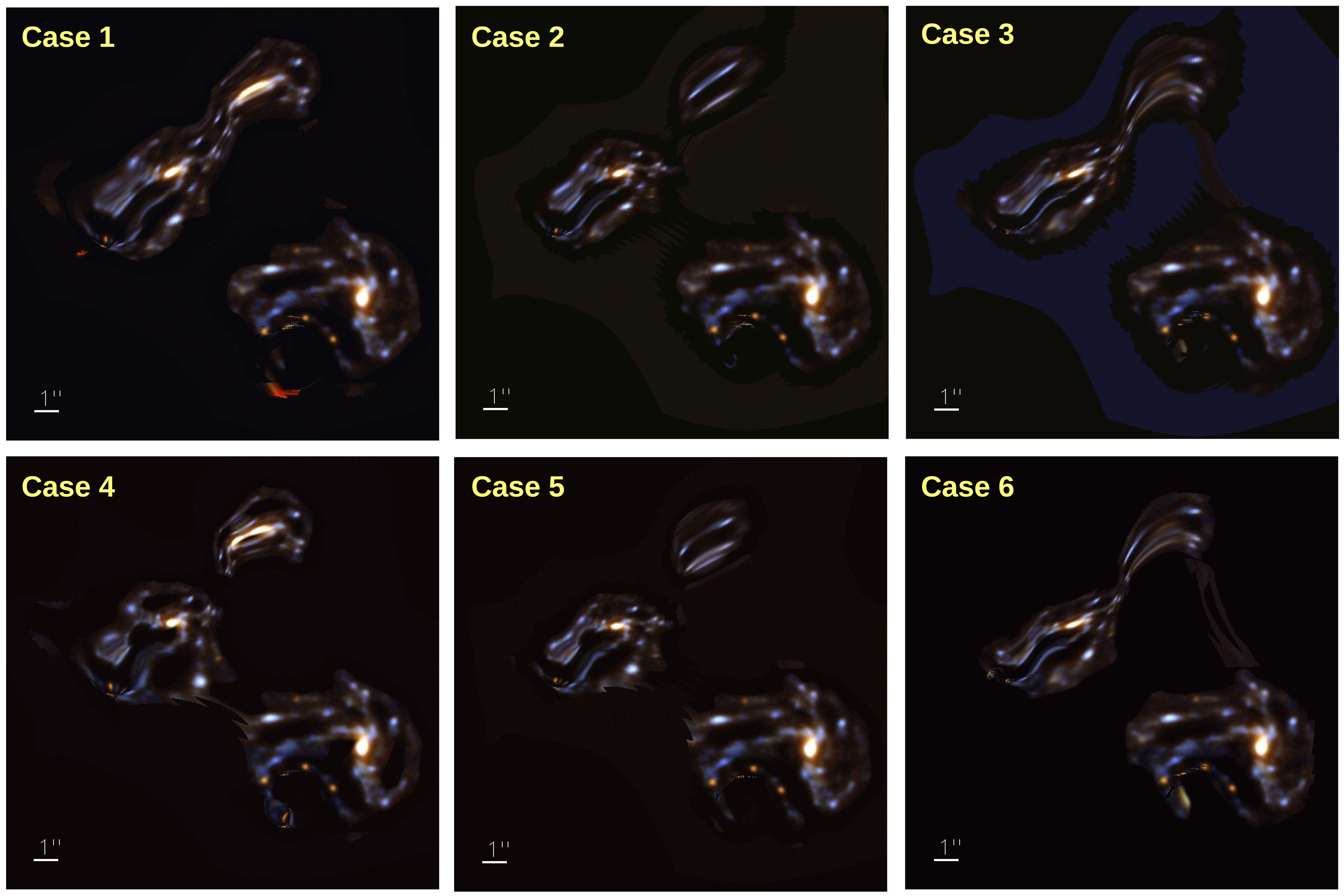} 
   \caption{Predicted images using the delensed 1.1 as a template of the source for the six different cases discussed in section \ref{sect_S6}. Note how models 1,3,4 and 6 predict and extra counterimage for the nucleus that is not observed. The scale and centre of the images are the same 
            as the left panel of Fig.  \ref{fig_Relens1a}.}
    \label{fig_Relens1a_6cases}  
\end{figure*}

\section{Results}\label{sect_S7}
\begin{figure}  
   \includegraphics[width=8.5cm]{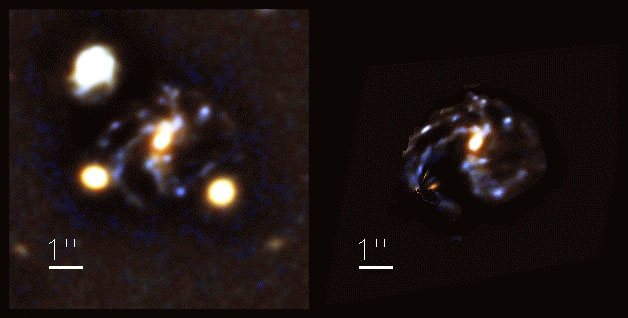} 
   \caption{Like in Fig. \ref{fig_Relens1a} but for the counterimage 1.3 in the north and for case 5. 
            The two {\it twin} elliptical galaxies in the south are included in our model.   
            The blue spiral galaxy in the north-east is 
            not included in our model and may still play a non-negligible role by stretching the deflection field in the direction 
            north-west.}
    \label{fig_Relens1b}  
\end{figure}

For each one of the six cases discussed in section \ref{sect_S6} we derive a solution. The minimization is stopped once the solution has converged to a stable point (typically between 50000 and 150000 iterations). 
In our earlier implementation of the code, such a large number of iterations would have produced solutions that predict unphysically small sources. This case is explicitly shown in figure \ref{fig_delened1c} (singular case), a situation that is avoided in our new implementation once the additional information regarding the source shape and size (for well resolved and extended arcs) is incorporated. 
We use the same number of iterations to derive the six solutions.  
A first comparison between the different solutions is made by contrasting the reconstructed shape of the original (delensed)  galaxy of system 1 in the source plane. 
Figure \ref{fig_delened1c} shows the predicted shape of the galaxy of system 1 in the source plane based on image 1.3. 
We find that changing the scale radius of the member galaxies has no significant effect on the deflection field around image 1.3. 
This is not surprising as 1.3 lies in a region of the cluster with no major member galaxies. 
Consequently, the delensed images of Cases 1,2, and 3 are virtually indistinguishable from those of Cases 4,5, and 6 respectively.
Comparing the different cases in Fig. \ref{fig_delened1c}, increasing the resolution of the grid seems to result in an elongation of the galaxy in the vertical direction. 
The right panel in the same figure, shows the corresponding solution for the case where no prior information about the shape of the galaxy is included and the algorithm reconstructs 
an unphysically small galaxy in the source plane. This is the solution that would have been derived with a very large number of iterations in our previous version of the code. {Although Fig. \ref{fig_delened1c} represents only the delensed version of image 1.3, using 1.1 renders similar results. The other counter images (1.2, 1.4 and 1.5) produce more distorted and/or partial reconstructions of the source.
  
A further assessment of the quality of the solution can be tested by comparing the observed and predicted counterimages in the image plane. We focus on system 1 as this is the most interesting one 
in terms of complexity and also due to its proximity to the centre of the cluster. Figure \ref{fig_Relens1a} shows the challenging cases of images 1.2, 1.4 and 1.5 (for the model 
corresponding to Case 5 described in the previous section). For this example, we use 1.1 as a template for the source that is delensed and relensed by our lens model to predict 
the other counterimages. The agreement between the observed and predicted images is in general very good with typical distances between observed and predicted features smaller than 1\arcsec. 

The prediction for all six models is shown in figure \ref{fig_Relens1a_6cases}. In general all models reproduce the observed image reasonably well.   
Image 1.4 is the most challenging and some significant deviations can be appreciated in some of the models. In particular, models 1,3,4 and 6 predict an additional 
counterimage for the nucleus that is not observed in the current data although it cannot be ruled out that the counterimage for the nucleus is lost in contrast with the central glare of the BCG. 
The upcoming deeper optical data from the HFF program (specially the UV band where the BCG will be relatively faint) will certainly help in testing for the existence of this possible additional image of the bulge of system 1.\footnote{The new HST UV data does not show any counterimage of the bulge in 1.4} 
Knot number 8 (see figure \ref{fig_systemID}) in image 1.2 seems to be reproduced better (closer to the BCG) in the case of the regular grid with the more extended (i.e larger scale radius) galaxies (case 1) suggesting that the mass 
distribution stretches and flattens in the direction connecting the BCG with the elliptical galaxy next to 1.2. Regarding the SN, no counterimage of the SN is expected around the BCG as shown in Fig. \ref{fig_TimeDelay} below.

The result for the counterimage 1.3 is shown in figure \ref{fig_Relens1b}  for case 5. As in the previous case, 1.1 is used to delens 
and relens the galaxy. The agreement is again very good, with the exception of the SN that appears in 4 locations while a perfect model would predict only one location}. This is due, in part, to the fact that the elliptical in between the SN is modeled 
as a spherical halo. We should not that the observed image 1.3 does not show the SN as it probably appeared at this position several years ago and was missed by the observations while the prediction above is based on an image that does contain the SN.

\begin{figure}  
   \includegraphics[width=8.5cm]{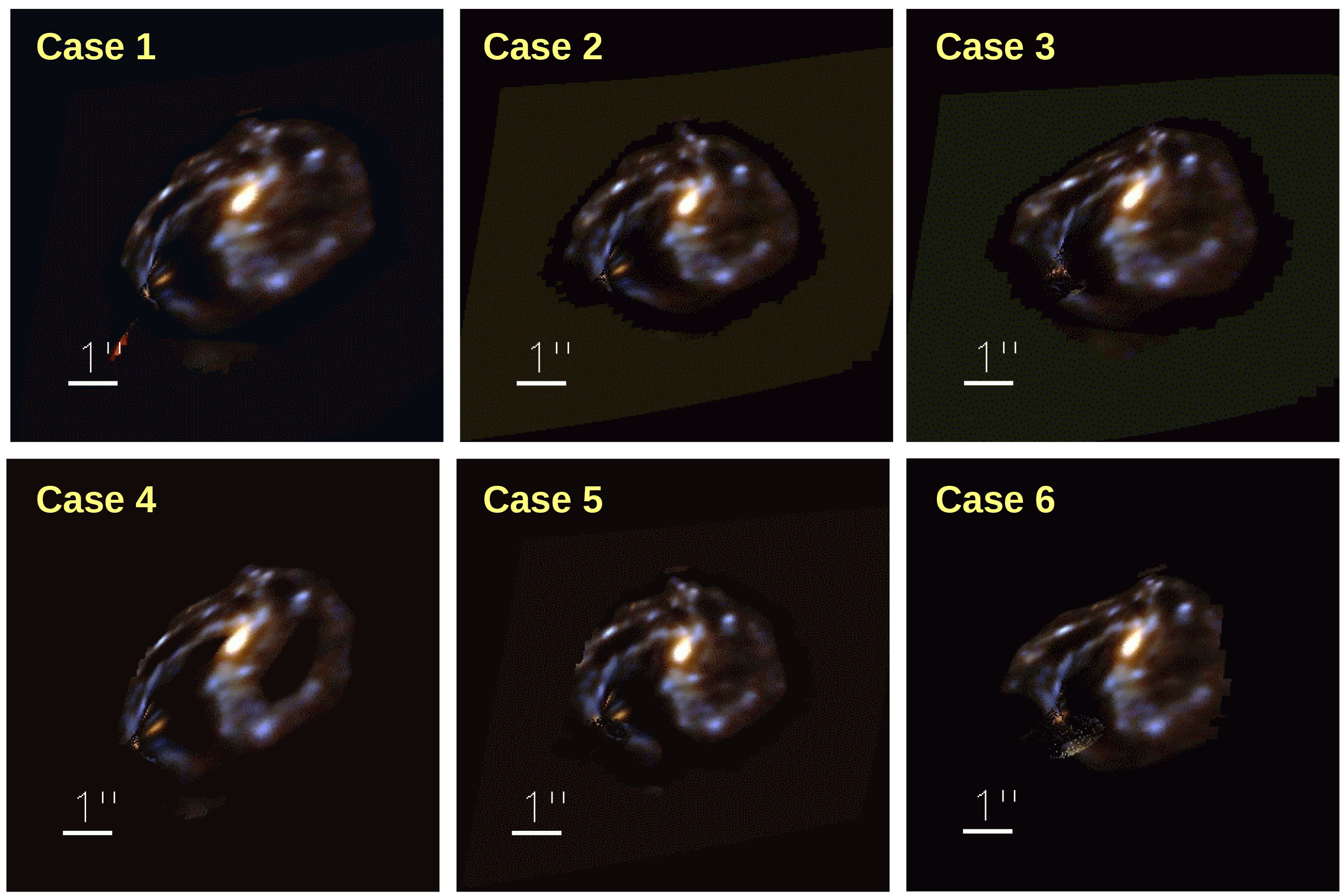} 
   \caption{Like figure \ref{fig_Relens1a_6cases} but for the predicted counterimage 1.3 in the northern part.}
    \label{fig_Relens1a_6cases_North}  
\end{figure}

Future improvements to the lens model will include a halo based on the elongation of the luminous matter \citep{Chen2015}. 
The predicted images 1.3 for the six models are shown in figure \ref{fig_Relens1a_6cases_North}. All images are centered on the same position as the left panel of 
figure  \ref{fig_Relens1b} and have the same scale. The agreement for all models is also excellent, with some models like cases 3 and 6 (high resolution grid) 
showing some distortion in the northern part of the galaxy. Cases 1 and 4 (regular grid) show an elongation in the diagonal direction. These distortions may be connected with the fact that 
the spiral galaxy in the north-east (possibly a cluster member) was not included in our set of cluster members as it is not included in the red-sequence. Due to the proximity of 
this spiral galaxy to image 1.3, a small distortion in the deflection field might be expected. The predictions from Cases 2 and 5 (intermediate resolution grid) seem to best reproduce image 1.3. 

The solutions show some sensitivity to the redshifts of systems with photo-z. In particular to images 4.3 and 8.3. Excluding images 4.3 and 8.3 from 
our analysis produces more smooth solutions and a rounder critical curve. These solutions (without 4.3 and 8.3) reproduce better the image pairs 4.1, 4.2 and 8.1, 
8.2 with a lower redshift. Imminent new optical data from the HFF should improve photo-z estimations and provide new systems that will help constrain the solution better. 
This will be explored in a future paper \citep{Treu2015}.

\subsection{Mass profile and mass distribution}

\begin{figure}  
 \centerline{ \includegraphics[width=9cm]{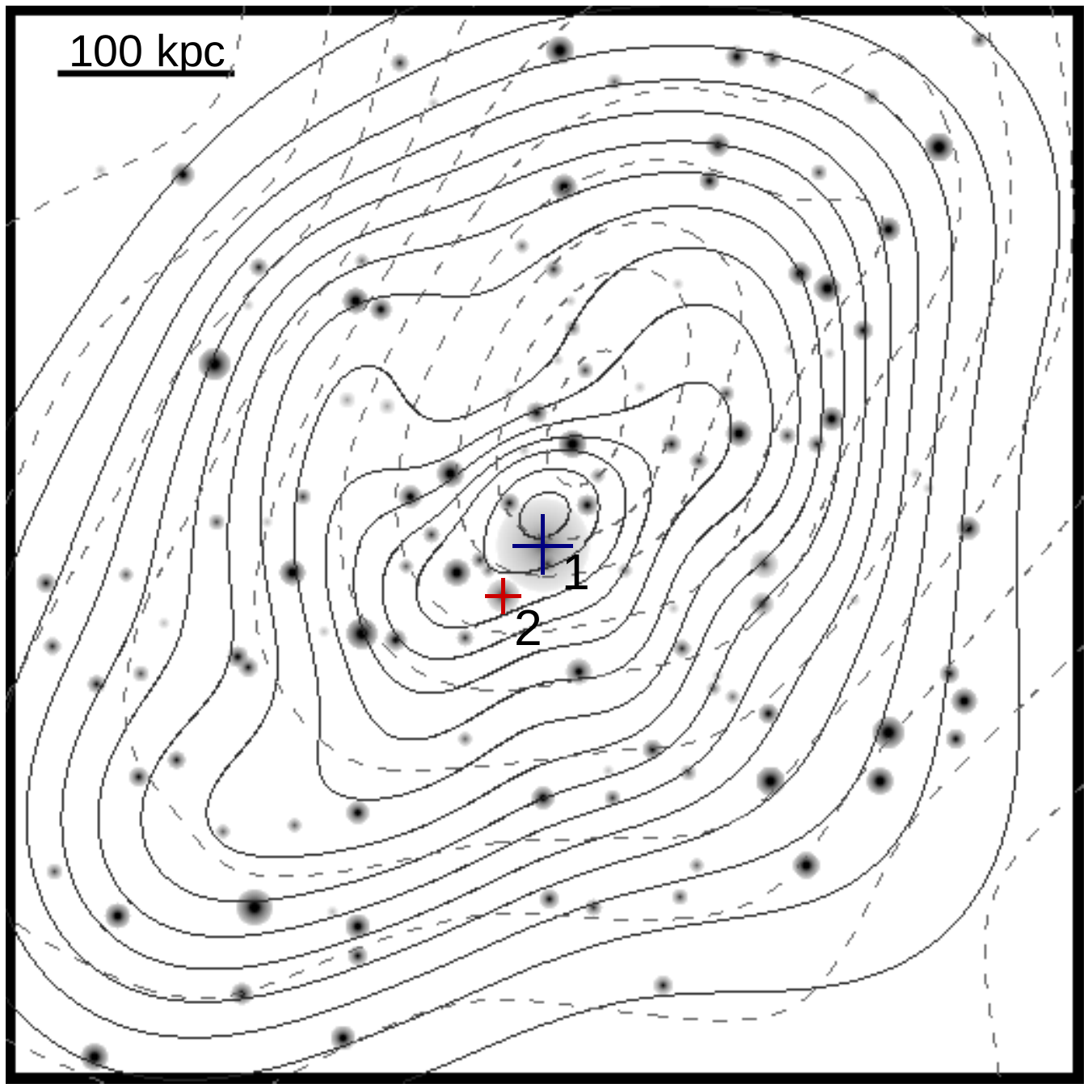}}  
   \caption{Contours of the grid component of the mass distribution for case 5 (solid lines) compared with the X-ray emission 
            as seen by {\it Chandra} (dashed lines). 
            The input galaxies used in our model (for cases 4,5 and 6) are shown for comparison. The galaxies defining layers 
            1 and 2 are marked with a big blue and small red cross respectively. The remaining galaxies form the layer 3 in our 
            model}. 
   \label{fig_MassChandra}  
\end{figure}

The 2-dimensional distribution of the soft component (grid) for the mass is shown in Fig. \ref{fig_MassChandra} and is compared with the position (and shape) 
of the input galaxies and with the X-ray emission from  {\it Chandra}. A few interesting conclusions can be derived from this plot. First, the peak of 
the soft component lies close to the position of the BCG. The small misalignment may be natural since the BCG may not be at rest with respect to the projected center of mass or this may be a consequence of our assumption of a spherical halo for 
the BCG while clearly the HST data suggests an elongation for this halo. The location of the peak and the elongation of the soft component around the BCG seems to correct 
for this wrong assumption by adding an elongation to the global mass distribution (or deflection field). 
Interestingly, this elongation points in the direction of the X-ray 
peak that is offset from the BCG by $\approx 50$ kpc. Also, in the direction of the X-ray peak there is another prominent member galaxy that was not fitted independently 
in our model. Better constraints around this massive galaxy derived from the imminent new HFF optical data will help constrain this galaxy in an independent way.  The offset between the X-ray and mass peaks confirms the disturbed nature of this cluster.

\begin{figure}  
 \centerline{ \includegraphics[width=10cm]{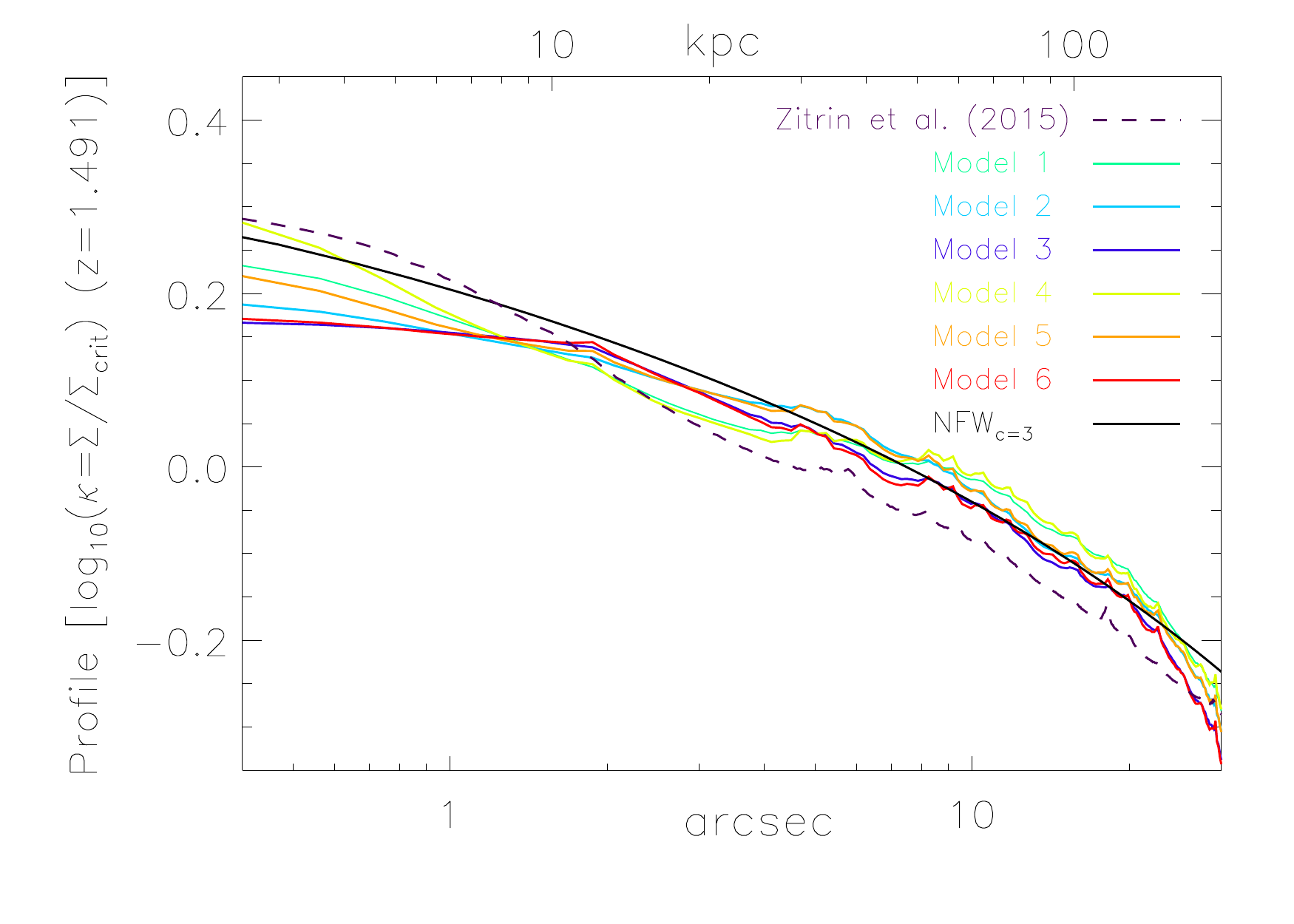}}  
   \caption{Mass profile in terms of the critical surface mass density (computed at $z=1.491$) 
            for the six solutions obtained with the regular grid  and with the 
            multiresolution grids. The dashed line is the recent estimate from Zitrin et al. (2015). The solid black line 
            is a projected NFW profile with concentration $C=3$ and $R_{200}=5$ Mpc.} 
   \label{fig_Profiles}  
\end{figure}

The mass profiles for the six models are shown in Fig. \ref{fig_Profiles}. A general good agreement is found between the six models. For comparison we show the profile 
derived recently by \cite{Zitrin2015} from the CLASH survey. A small shift is observed between the parametric solution of \cite{Zitrin2015} and our free-form solution. The origin of this 
small discrepancy may be the fact that for systems 4 and 8 we use a redshift that is different than the one assumed in \cite{Zitrin2015}.  
In contrast with our previous work on HFF clusters (A2744, MACS0516 and MACS0717), 
no plateau is observed in the profile beyond the outer radius of the central galaxy.   We find a good agreement between our derived profile and a low concentration  NFW profile expected for massive clusters. 
According to results from simulations \citep{Meneghetti2014}, and recent observations on clusters  \citep{Merten2014}, massive galaxy clusters are well reproduced by NFW 
profiles with relatively low values of the concentration parameter, $C \approx 3-4$, with somewhat larger values are derived for well defined relaxed 
clusters from the CLASH program \citep{Umetsu2014,Zitrin2015}. In the particular case of MACS1149, we find that a NFW profile with a concentration $C=3$ produces a good match to the observed projected profile, consistent with the dynamical state of this cluster which from our comparison of the gas and dark matter distribution (Fig. \ref{fig_MassChandra}) is evidently not suffering a first core passage of a major merger, but is not yet well relaxed.

\subsection{Time delays}\label{sect_timedelay}

Here we use our free-form models to obtain the 2D-time delay surface (and uncertainty) for this cluster. 
Image 1.1 hosts a quadruply lensed SN (and named in honour of  Refsdal for his pioneering interest in this regard, \cite{Refsdal1964}). 
Several authors have predicted the time delays between the 4 SN counterimages in 1.1 
and the predicted position of the SN in images 1.2 and 1.3. \cite{Oguri2015} predicts the SN in 1.3 
appeared 17 years ago while one in 1.2 will appear in 1-3 years. 
In \cite{Sharon2015}, the authors predict the SN in 1.2 will appear in $\approx$ 0.65 years (around early July 2015 with $\approx$ 1 month uncertainty) after the SN is 
observed in 1.1. The same model predicts the SN in 1.3 occurred  $\approx$ 11.6 years before the SN in 1.1 was observed (with $\approx$ 1 year uncertainty). 
The suite of models produced by the method of \cite{Zitrin2009a,Zitrin2015} as presented in \cite{Kelly2015} also predicts time delays consistent with these predictions although with larger error bars in part due to the exploration of a wider range of models and the addition of uncertainties in the photometric redshifts. In contrast, we restrict the computation of our uncertainties to just the six models presented above and no errors are adopted for the photometric redshifts so our error bar is likely to be underestimated. 

We compute time delays from our 6 models and estimate the mean and dispersion from these 6 models. 
The time delay is defined as
\begin{equation}
t(\theta) = \frac{1+z_d}{c}\frac{D_dD_s}{D_{ds}}\left[ \frac{1}{2}(\theta - \beta)^2 - \psi(\theta) \right].
\label{eq_timedelay}
\end{equation}

\begin{figure}    
 {\includegraphics[width=8.5cm]{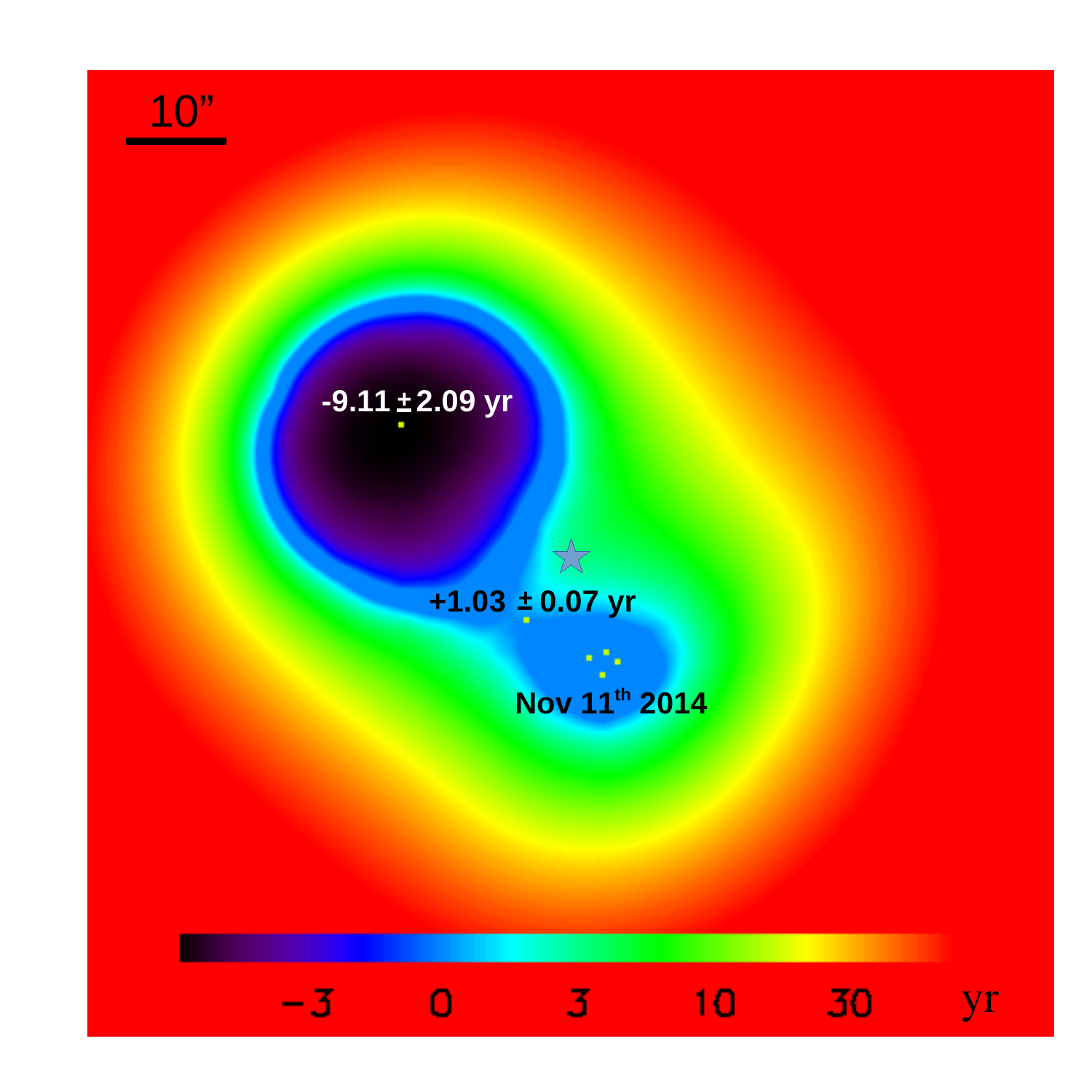}}
   \caption{Time delay surface computed with respect to the quadruply lensed SN. The yellow dots mark the location of the 4 SN images and the predicted positions of 
            the SN in two counterimages. The time delays for the additional 
            counterimages lie $\approx$ 9 years in the past for the counterimage 1.3 and $\approx$ 1 year in the future for the counterimage 1.2. The star marks the position of the cluster BCG.}
   \label{fig_TimeDelay}  
\end{figure}

Figure \ref{fig_TimeDelay} shows the average time delay from our models. The counterimage in 1.2. lies in the 
future by approximately 1 year with respect to the observed SN. If the model prediction is correct, the SN should appear again 
by the end of 2015. The counterimage of the SN at image 1.3 is predicted to have occurred approximately 9 years ago.  These predictions 
are similar to those derived by \cite{Sharon2015}. Also. a visual comparison of our Fig.  \ref{fig_TimeDelay} with 
the bottom-left panel of Fig. 4 in \cite{Sharon2015} reveals a similar structure in the 2-dimensional distribution of the time delay. 
The agreement between our free-form prediction and their parametric predictions favours strongly a window around July-December 2015 to observe the 
SN in image 1.2 at RA=11:49:36.025, DEC=+22:23:48.11 (J2000).

A measurement of the time delay in the near future can be used to impose tight constraints in the lens model. 
Small differences between models result in small changes in the balance between the terms $(\theta - \beta)^2$ 
and $\psi(\theta)$ which are later magnified by the factor  $\frac{1+z_d}{c}\frac{D_dD_s}{D_{ds}}$. 
This factor typically takes values of $\sim 1$ Gyr. A discrepancy in the time delay prediction of one year between models is possible with 
small changes of order $10^{-9}$ in the difference shown in brackets in Eq. \ref{eq_timedelay}. 
Note that these models do not predict the time of explosion of the SN, which is estimated empirically to be about ~3 weeks prior to the November 
2014 NIR observations in which it was discovered \citep{Kelly2015}, so that when predicting the date of the future SN explosion anticipated for image 1.2, 
we must subtract off about ~3 weeks.
We estimate then that the SN will reappear around November the 1$^{st}$, and with an uncertainty of 25 days. 
If the SN is in fact observed in the near future in image 1.2, this observation could be used to improve the constraints in the lens 
model and perhaps even cosmological parameters like the Hubble constant. For the current paper, we have adopted the value $h=0.7$ for the Hubble constant. 
While the strong-lensing constraints are not sensitive to h (due to the degeneracy in $h$ between the geometric factor and the cluster mass), the time delay 
exhibits a different dependency with the Hubble constant. Accurate estimations of the mass based on strong lensing constraints can be used to derive precise 
predictions of the time delay that scale as $h^{-1}$ and when contrasted with measurements of these time delays, derive a constraint on $h$ \citep{Oguri2007}. 
 
In the particular case of WSLAP+, time delay constraints can be easily incorporated if one makes the approximation that the change in position 
of the background source is small (in relation to the typical deflection angles) when time delays constraints are incorporated in the reconstruction. 
In this case, the unknown variable $\beta$ in the quadratic term $(\theta - \beta)^2$ can be expressed as a fixed term, $\beta_0$, 
plus a small perturbation, $\delta\beta$.

\begin{eqnarray}
(\theta - \beta)^2 & = &(\theta - (\beta_0 + \delta\beta))^2 \nonumber \\
                   & = & (\theta - \beta_0)^2 +  \delta\beta^2 - 2\theta\times\delta\beta \nonumber \\ 
                & \approx & C - 2\theta\times\delta\beta 
\end{eqnarray}

where $\beta_0$ is the source position as inferred from the arc positions in the standard strong lensing analysis, $\delta\beta$ is the offset 
(with respect to $\beta_0$) of the new source position when time delays are included, C is a known variable (constant) that can be pre-computed and we 
make the approximation that the term $\delta\beta^2$ is much smaller than the other terms so it can be neglected. This assumption is good if in fact the 
offset between the new source position and that inferred from the strong lensing constraints is indeed small when compared with the typical 
deflection angles. When computing $\delta\beta$ for our 6 solutions, we find that the relative change between the six models is indeed very small 
(less than one arcsecond). 
Hence, if time delays are able to discern between different models (like the ones used in this work), the approximation that $\delta\beta^2$  is very 
small is valid. Under this approximation, Eq. \ref{eq_timedelay} can be linearised in the unknown variables (mass and source position) 
and solved using the same fast optimization algorithm (system of linear equations). Fixing $\beta_0$ requires solving 
the problem in an iterative way where each time the value of $\beta_0$ is updated. The convergence of the algorithm when time delays are included 
will be tested in a future work. 

\section{Conclusions}\label{sect_S8}
We apply our improved lensing reconstruction algorithm to the HFF cluster MACS1149. The new enhancements  in the code to incorporate the internal structure of the large spiral galaxy images, in which the Refsdal SN was found, result in solutions that are more stable and precise. 
The increased precision in the solutions allows us to reproduce the observed images with unprecedented detail for a free-form method. Our best fitting mass distribution shows a somewhat disturbed but single cluster well fitted radially by an NFW profile with a low concentration value, $C=3$, typical for large unrelaxed clusters \citep{Neto2007}. We confirm an offset between the X-ray and mass maxima, that is similar to the offset observed between the centre of the cluster BCG and the X-ray peak. The peak of the diffuse mass distribution extends towards the position of the 
X-ray peak suggesting that the lensing data may be sensitive to the X-ray plasma, a possibility already suggested by our earlier HFF work on A2744 and 
MACS0416 \citep{Lam2014,Diego2015}. 

Our  improved model allows us to compute precise time delays 
for the observed Refsdal SN. Our results are in agreement with previous estimates and places the future occurrence of the 
SN somewhere between early October 2015 and early January 2016 (assuming a value for the Hubble constant of h=0.7). 
A significant deviation from this prediction will result in changes in the lens model and/or the cosmological model (in particular h). The inclusion of 
a future observation of this time delay can be easily incorporated in our reconstruction algorithm and will be exploited in future work. 
The planned deep optical observations of this cluster as part of the 
ongoing HFF will reveal new multiply lensed images that will help in constraining better this cluster, 
in particular the halos of the perturbing central member galaxies including the BCG, for which currently the data are still very ambiguous.

\section{Acknowledgments}  
This work is based on observations made with the NASA/ESA {\it Hubble Space Telescope} and operated by the Association of Universities for Research in Astronomy, Inc. 
under NASA contract NAS 5-2655. 
Part of the data for this study is retrieved from the Mikulski Archive for Space Telescope (MAST).
The authors would like to thank the HFF team for making this spectacular data set promptly available to the community.
The scientific results reported in this article are based in part on data obtained from the Chandra Data 
Archive 
\footnote{ivo://ADS/Sa.CXO\#obs/1656},\footnote{ivo://ADS/Sa.CXO\#obs/3589},\footnote{ivo://ADS/Sa.CXO\#obs/16238} 
\footnote{ivo://ADS/Sa.CXO\#obs/16239},\footnote{ivo://ADS/Sa.CXO\#obs/17595},\footnote{ivo://ADS/Sa.CXO\#obs/17596} 
\footnote{ivo://ADS/Sa.CXO\#obs/16306},\footnote{ivo://ADS/Sa.CXO\#obs/16582}.
We would like to thank Harald Ebeling for making  the code {\small ASMOOTH} \citep{Ebeling2006} available. 
T.J.B. thanks the University of Hong Kong for generous hospitality. J.M.D acknowledges support of the consolider project 
CSD2010-00064 and AYA2012-39475-C02-01 funded by the Ministerio de Economia y Competitividad. 
AZ was provided by NASA through Hubble Fellowship grant \#HST-HF2-51334.001-A awarded by STScI.
We are grateful the anonymous referee for his/her valuable suggestions and comments that have helped us improve the contents of this paper. The authors would like also to thank Elizabeth E. Brait for assistance with some of the graphical work of this paper.
  
\label{lastpage}
\bibliographystyle{mn2e}
\bibliography{MyBiblio} 

\end{document}